\documentclass[
reprint,
superscriptaddress,
 amsmath,amssymb,
aps,
prb,
floatfix,
]{revtex4-2}

\usepackage{graphicx}
\usepackage{dcolumn}
\usepackage{bm}
\usepackage{hyperref}
\usepackage{xcolor}
\usepackage{soul}

\begin{document}

\preprint{APS/123-QED}

\title{Probing Fermi surface parity with spin resolved transverse magnetic focussing}

\author{M. J. Rendell}
    \thanks{M. J. Rendell and S. D. Liles contributed equally to this work}
\author{S. D. Liles}
\author{S. Bladwell}
\author{A. Srinivasan}
\affiliation{
School of Physics, University of New South Wales, Sydney, NSW 2052, Australia
}
\author{O. Klochan}
\affiliation{
University of New South Wales Canberra, Canberra, ACT 2600, Australia
}
\author{I. Farrer}
\affiliation{
Department of Electronic and Electrical Engineering, University of Sheffield, Sheffield, S1 3JD, UK
}
\author{D. A. Ritchie}
\affiliation{
Cavendish Laboratory, University of Cambridge, Cambridge, CB3 0HE, UK
}
\author{O. P. Sushkov}
\author{A. R. Hamilton}
\affiliation{
School of Physics, University of New South Wales, Sydney, NSW 2052, Australia
}
    \email{matthew.rendell@unsw.edu.au, alex.hamilton@unsw.edu.au}
\date{\today}

\begin{abstract}
Measurements of the Fermi surface are a fundamental technique for determining the electrical and magnetic properties of solids. In 2D systems, the area and diameter of the Fermi surface is typically measured using Shubnikov-de Haas oscillations and commensurability oscillations respectively. However, these techniques are unable to detect changes in the parity of the Fermi surface (i.e. when E(+k) $\neq$ E(-k)). Here, we show that transverse magnetic focussing can be used to detect such changes, because focussing only measures a well defined section of the Fermi surface and does not average over +k and -k. Furthermore, our results show that focussing is an order of magnitude more sensitive to changes in the Fermi surface than other 2D techniques. While we investigate a specific Fermi surface shift in this work, focussing could be used to investigate similar Fermi surface changes in other 2D systems.

\end{abstract}
\maketitle

\begin{figure}
    \centering
    \includegraphics[width=\linewidth]{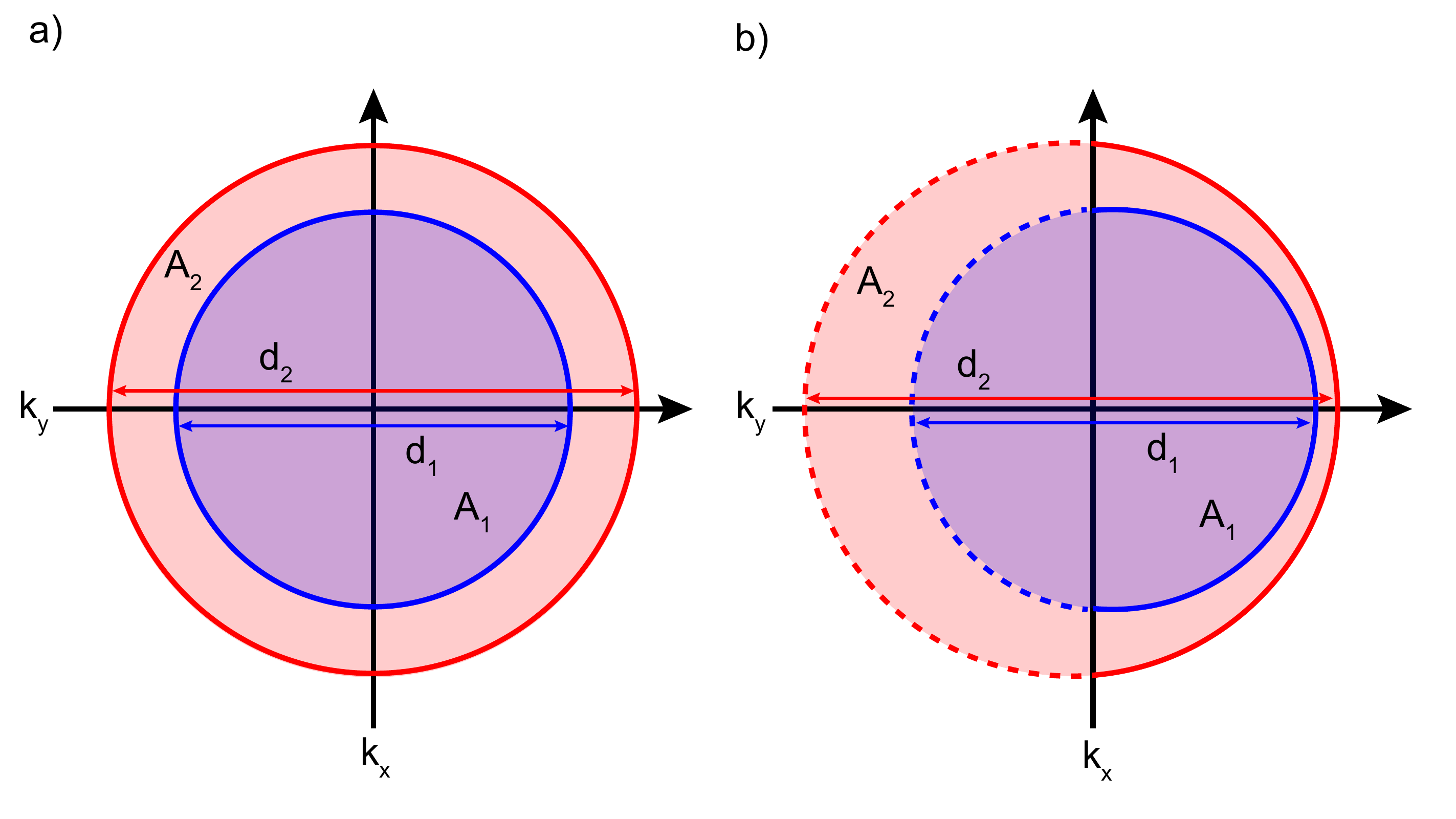}
    \caption{\textbf{Comparison of Fermi surface measurement techniques. a)} A typical 2D Fermi surface which has been spin split by a spin-orbit interaction. This spin-splitting can be measured using Shubnikov-de Haas oscillations (Fermi surface area A$_1$ and A$_2$) and commensurability oscillations (Fermi surface diameter d$_1$ and d$_2$). \textbf{b)} As in a) with the addition of an in-plane magnetic field which causes a relative shift in the Fermi surface. Magnetic focussing only measures one arc of the Fermi surface (indicated by the solid section), and is able to detect the Fermi surface shift despite the Fermi surface area and diameter remaining unchanged.}
    \label{fig:TechniqueComparison}
\end{figure}

Techniques for measuring the Fermi surface have existed since the early 1930s \cite{de_haas_dependence_1930, schubnikow_new_1930}, and are a powerful way to probe electronic and magnetic properties of metals, semiconductors, superconductors and heavy fermion compounds \cite{lonzarich_magnetic_1988, hussey_coherent_2003, doiron-leyraud_quantum_2007, danzenbacher_insight_2011}. In 2D systems, the area of the Fermi surface is typically measured using Shubnikov-de Haas oscillations (A$_1$ and A$_2$ in Figure~\ref{fig:TechniqueComparison} a), while the diameter can be measured using commensurability oscillations (d$_1$ and d$_2$). \cite{soule_study_1964, askenazy_oscillations_1969, shaw_shubnikov-haas_1970, brosh_probing_1996, skuras_anisotropic_1997, kamburov_anisotropic_2012}. These techniques allow for the size and asymmetry of the Fermi surface to be determined experimentally, however they are not able to resolve a change in the parity of the Fermi surface (i.e. when E(+k) $\neq$ E(-k)) as this is averaged out over a full loop of the Fermi surface. Unfortunately such changes in parity are quite common, for example due to trigonal warping or the combination of spin-orbit interactions and an in-plane magnetic field \cite{predin_trigonal_2016}. Figure~\ref{fig:TechniqueComparison} b) shows an example of such a Fermi surface, where a combination of a Rashba spin-orbit interaction and an in-plane magnetic field removes the symmetry around $k_x=0$. In this situation the area and diameter of the Fermi surface do not change so there is no effect on Shubnikov-de Haas or commensurability oscillations.

A lesser used technique for measuring Fermi surfaces is transverse magnetic focussing. Originally used to probe the Fermi surface of metals \cite{sharvin_possible_1965, tsoi_focusing_1974, tsoi_transverse_1996}, magnetic focussing has also revealed spin and charge dynamics including branched electron flow, small-angle scattering, spin separation and spin precession \cite{aidala_imaging_2007, bladwell_interference_2018, gupta_precision_2021, rendell_spin_2023, rokhinson_spin_2004, lo_controlled_2017}. Because magnetic focussing only measures a well defined section of the Fermi surface (designated by the solid arcs in Figure~\ref{fig:TechniqueComparison} b) \cite{van_houten_coherent_1989, heremans_observation_1992, heremans_transverse_1994, taychatanapat_electrically_2013, lee_ballistic_2016, bachmann_super-geometric_2019, berdyugin_minibands_2020, rao_ballistic_2023, balduini_probing_2024}, it is sensitive to Fermi surface changes that average out over a full orbit.

Here we show how this unique feature of transverse magnetic focussing can be used to detect changes in the parity of the Fermi surface. As a concrete example, we use the simple case of a Rashba spin-orbit interaction combined with an in-plane magnetic field. We show that the focussing signal is able to detect a change in Fermi surface parity caused by the direction of the in-plane field. We also demonstrate the high sensitivity of this technique, and are able to detect changes to the Fermi surface caused by fields as small as 0.1T.

\begin{figure}
	\includegraphics[width=0.45\textwidth]{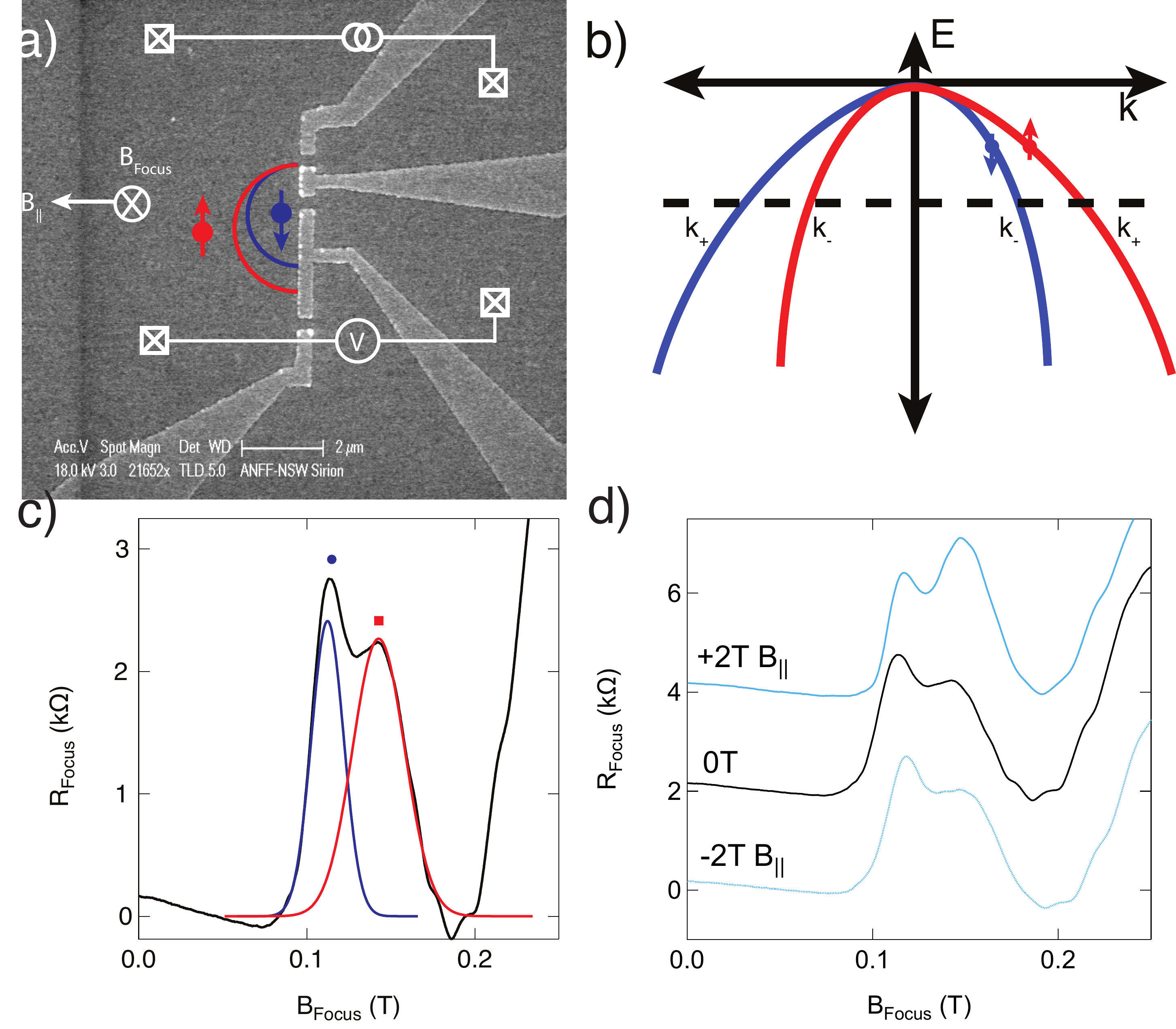}
    \caption{\textbf{Spin-resolved focussing with in-plane magnetic fields a)} SEM of the focussing sample. The overlay shows the orientation of the in-plane (B$_{\parallel})$ and out-of-plane (B$_{Focus}$) magnetic fields, as well as the electrical setup for measurements. Red and blue semicircles indicate the spin-split focussing trajectories. \textbf{b)} The first 2D subband for holes. A Rashba spin-orbit interaction causes a spin-splitting of the subband resulting in two different momenta (k$_+$ and k$_-$) at the Fermi energy (horizontal dashed line). This difference in momentum results in different focussing trajectories for each spin. \textbf{c)} The spin split first focussing peak resulting from the two spin-dependent focussing trajectories (labelled with blue circle and red square). By fitting a double Gaussian to the first focussing peak the amplitude of both peaks can be extracted. \textbf{d)} Spin split focussing with an in-plane magnetic field. The change in peak amplitude is not symmetric when the direction of the in-plane field is reversed.}
    \label{fig:FocussingDiagram}
\end{figure}

The magnetic focussing sample is fabricated on a GaAs/Al$_{0.33}$Ga$_{0.67}$As heterostructure with a 15nm GaAs quantum well confining the 2D hole gas (2DHG) 85nm below the surface. The 2DHG is induced in accumulation mode (no doping) by applying a negative voltage to an overall top gate. The use of an undoped heterostructure avoids device instability in the gate defined nanostructures which create the focussing geometry \cite{see_impact_2012, srinivasan_improving_2020}. Figure~\ref{fig:FocussingDiagram} a) shows an SEM of the sample with lithographic split gates used to define the focussing geometry. The overlay indicates the electrical measurement setup and the orientation of the in-plane magnetic field. To perform focussing, a constant current of holes (I$_\text{SD}$ = 5nA) is injected through a quantum point contact (QPC), and the resulting focussing signal is measured as a voltage across a second QPC. This is performed as a four terminal measurement with a pair of lock-in amplifiers at low frequency (17 Hz). When the perpendicular magnetic field (B$_\text{Focus}$) is such that the focussing diameter (d$_\text{Focus}$) is equal to the spacing between QPCs, a peak is observed in the focussing voltage. These peaks occur when the magnetic field is an integer multiple of \cite{van_houten_coherent_1989}
\begin{equation*}
    B_\text{Focus} = \frac{2 \hbar k_F}{e d_\text{Focus}}
\end{equation*}
Where k$_F$ is the Fermi momentum. All measurements in this work use a focussing diameter d$_\text{Focus}$ = 800nm at a 2D density of n$_\text{2D}$ = 1.89x10$^{11}$ cm$^{-2}$, (V$_\text{TG}$ = -1.35V) with a mobility of 760 000 cm$^{-2}$ V s and mean free path of 5.3$\mu$m. The QPCs have lithographic dimensions of 300x300nm and are biased to G=$2e^2/h$ to inject and detect both spin polarisations. The sample is measured in a He dilution system with a 9/5/1 T vector magnet at base temperature (30mK). In-plane and out-of-plane fields are also measured using Hall sensors on the sample probe to correct for any magnet hysteresis.

In GaAs hole systems, spin-orbit interactions have a significant effect on magnetic focussing measurements. The Rashba spin-orbit interaction causes a spin splitting of the first 2D subband which results in a different momentum for each spin. Figure~\ref{fig:FocussingDiagram} b) shows the spin-split first 2D subband for holes in GaAs, with the red and blue bands representing the different spins. The splitting of the 2D subband results in a difference of momentum between the spins (k$_+$ and k$_-$) at the Fermi energy (horizontal dashed line). The different momentum creates a different focussing trajectory for each spin, which splits the first focussing peak \cite{rokhinson_spin_2004}. The splitting of the 2D subband can also create a different scattering rate for each spin, as it changes the slope of each subband and hence the velocity. The relatively symmetric quantum well heterostructure used in this work allows for visible spin splitting while also giving a similar scattering rate for both spin states \cite{rendell_spin_2023}.

Figure ~\ref{fig:FocussingDiagram} c) shows focussing with no in-plane magnetic field. A double peak is observed consistent with spin-resolved focussing, with additional small oscillations due to the Shubnikov-de Haas effect and path interference \cite{van_houten_coherent_1989, bladwell_interference_2018}. Above B$_\text{Focus}$ = 0.2T the resistance increases due to the onset of the second classical focussing peak. A double Gaussian fit to the split peak (red and blue peaks in Fig.~\ref{fig:FocussingDiagram} c) shows that both peaks have a similar amplitude in the absence of an in-plane magnetic field. The higher field (red square) peak is also broader due to the previously mentioned path interference and a difference in scattering \cite{bladwell_interference_2018, rendell_spin_2023}. Figure~\ref{fig:FocussingDiagram} d) shows the first focussing peak with a magnetic field applied in-plane parallel to the QPC current direction (B$_{\parallel}$). With B$_{\parallel}$ = +2T (top trace - blue), a clear change in the peak amplitude is observed compared to the peaks with no in-plane field (black centre trace). This change in peak amplitude would typically be interpreted as a change in the spin polarisation \cite{rokhinson_spin_2004}. However, when  the direction of B$_{\parallel}$ is reversed to -2T (bottom trace in Fig.~\ref{fig:FocussingDiagram} d) the change in peak amplitude is not symmetric. This is not consistent with a change in spin polarisation as the Zeeman splitting should be the same for $\pm $B$_{\parallel}$.

\begin{figure}
    \centering
    \includegraphics[width=0.5\textwidth]{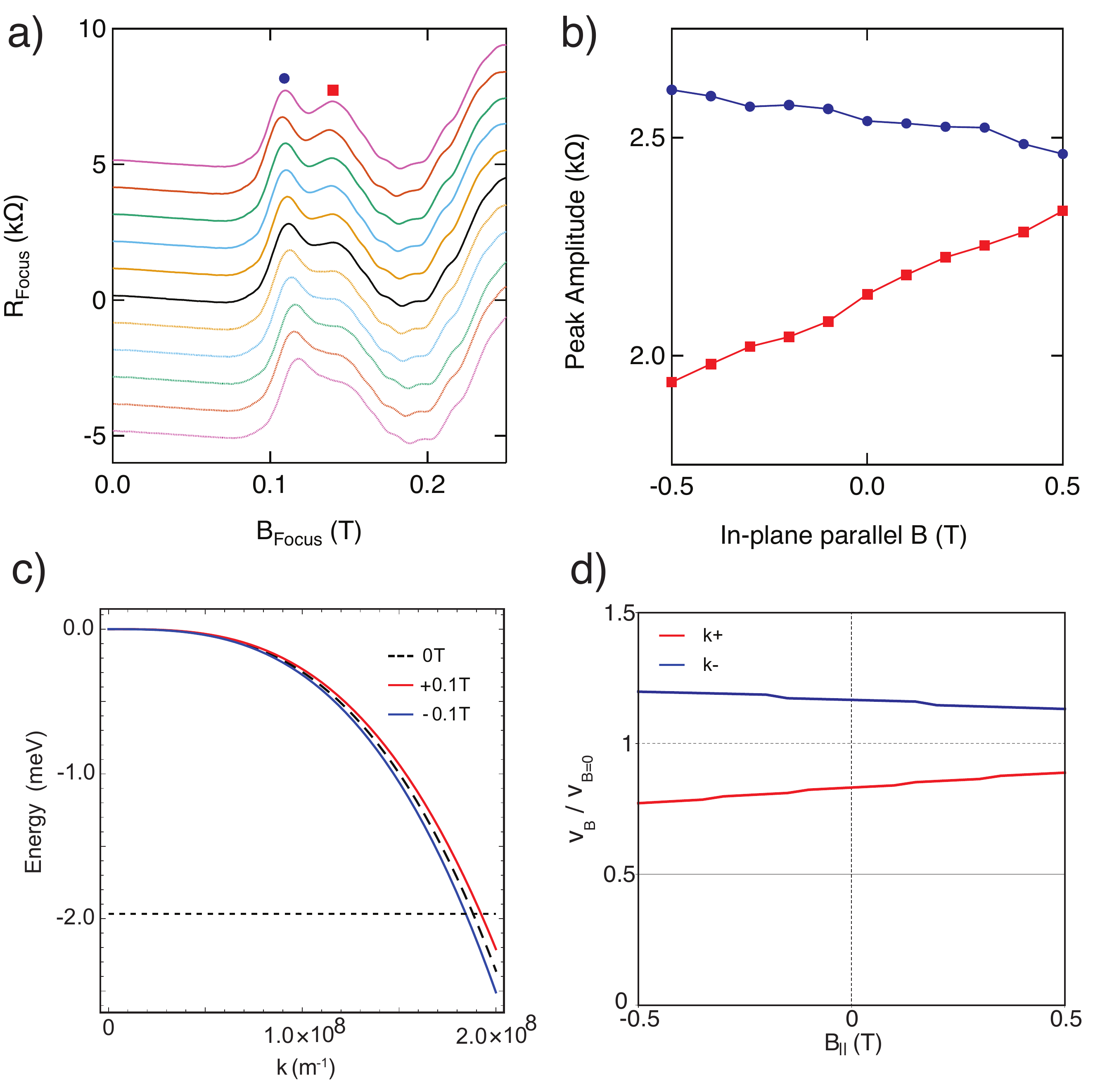}
    \caption{\textbf{Focussing with small B in-plane parallel to the QPC orientation.} \textbf{a)} Shows the focussing signal for different in-plane fields of $\pm$ 0.5 T in 0.1 T steps. Central black trace is for zero field, top solid traces are for field parallel to the QPC current and bottom dashed traces are for field antiparallel. Data has been vertically offset for clarity. \textbf{b)} The amplitude of each spin peak from a double Gaussian fit to the data in a). \textbf{c)} The calculated energy dispersion of one of the spin-split 2D hole subbands for B$_{\parallel}$ = +0.1T, 0T and -0.1T using Eq~\ref{eq:Hamiltonian}. The dashed horizontal line is the Fermi energy. \textbf{d)} The change in velocity of the spin subbands as a function of B$_{\parallel}$. The calculated velocity of each spin state (v$_B$) is normalised by the average velocity of the two spins at B$_{\parallel}=0$ (v$_{\text{B}=0}$).}
    \label{fig:SmallB}
\end{figure}

To rule out any Zeeman or spin polarisation effects, we next measure the change in peak amplitude with small in-plane magnetic fields. Figure~\ref{fig:SmallB} a) shows the results of focussing with small B$_{\parallel}$ applied in 0.1T increments up to $\pm$0.5T. A double Gaussian is fitted to each peak and the amplitude as a function of B$_{\parallel}$ is plotted in Fig.~\ref{fig:SmallB} b). A change in amplitude of the spin peaks is observed for fields as small as $\pm$0.1T, far too small to be caused by a change in spin polarisation. Even at 0.5T the change in peak amplitude ($\sim$ 10\%) is significantly larger than the Zeeman energy ($\sim$1\% of E$_\text{F}$) and therefore is too small to be a Zeeman effect. Instead, we consider a shift in the Rashba spin-splitting of the 2D subbands caused by B$_{\parallel}$.

The Hamiltonian for the 2D subbands is of the form
\begin{multline}\label{eq:Hamiltonian}
    \mathcal{H} = \frac{\mathbf{p^2}}{2 m^*} + \frac{i \alpha}{2} (\sigma_+ p^3_- - \sigma_- p_+^3) + \frac{g_1 \mu_B}{2} (B_+ p^2_+ \sigma_- + B_- p^2_- \sigma_+) \\
    + \frac{g_2 \mu_B}{2} (B_- p^4_+ \sigma_- + B_+ p^4_- \sigma_+)
\end{multline}
where $\frac{i \alpha}{2} (\sigma_+ p^3_- - \sigma_- p_+^3)$ is the Rashba spin-orbit term, $\frac{g_1 \mu_B}{2} (B_+ p^2_+ \sigma_- + B_- p^2_- \sigma_+) + \frac{g_2 \mu_B}{2} (B_- p^4_+ \sigma_- + B_+ p^4_- \sigma_+)$ are the Zeeman terms due to the in-plane magnetic field and $B_{\pm} = B_x \pm iB_y$. $B_{\parallel}$ causes a small shift in the 2D spin subbands, which is in opposite directions for the two subbands. Figure~\ref{fig:SmallB} c) shows the calculated energy dispersion of one of the 2D hole subbands using Eq.~\ref{eq:Hamiltonian}. There is a small shift in the subband dispersion, which is not the same for $\pm \text{B}_{\parallel}$. This shift is too small to cause a measurable change in the location of the magnetic focussing peaks, however it will still cause a change in the subband curvature \cite{bladwell_magnetic_2015, bladwell_interference_2018, marcellina_electrical_2018}. The change in subband curvature causes the Fermi velocity (v$_\text{F}$) to change along the hole trajectory. v$_\text{F}$ is linked to the scattering rate of each spin, since holes that travel slower will have more time to scatter (and vice versa). The focussing peak amplitude is exponentially sensitive to v$_\text{F}$: 
\begin{equation}
    R_{Focus} \propto e^{-\pi d/(2 v_\text{F} \tau_\text{Focus})} 
\end{equation}
where $d$ is the focussing diameter and v$_\text{F} \tau_\text{Focus}$ is the characteristic focussing scattering length \cite{heremans_observation_1992, rendell_transverse_2015, rendell_spin_2023}. Figure~\ref{fig:SmallB} d) shows the calculated change in v$_\text{F}$ for small B$_{\parallel}$. This matches the trend in focussing peak amplitude shown in Fig.~\ref{fig:SmallB} b) (see supplementary info for further comparison \cite{rendell_supp_info}).  This change in focussing peak amplitude for small B$_\parallel$ demonstrates the high sensitivity of focussing to small changes in the Fermi surface.

\begin{figure*}
	\includegraphics[width=\textwidth]{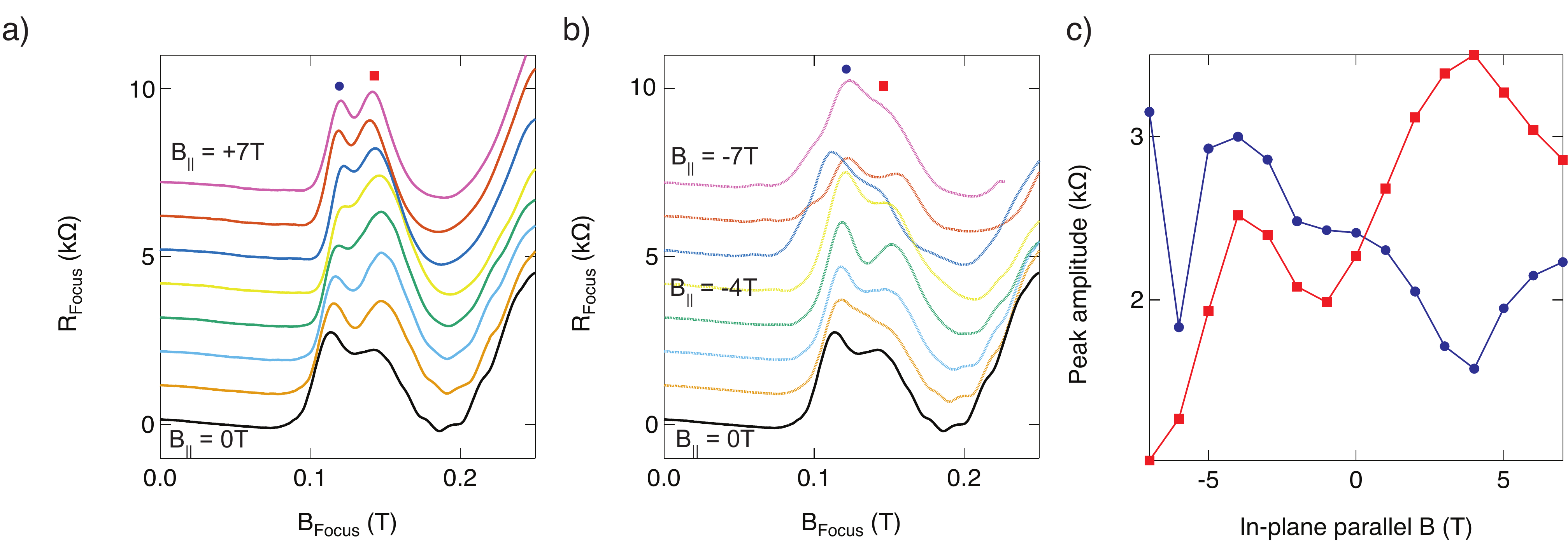}
    \caption{\textbf{Focussing with a large B in-plane parallel to the QPC current.} \textbf{a)} The focussing signal with an in-plane field applied along the QPC up to +7 T in 1 T steps. The black trace is for zero in-plane field. Each trace has been vertically offset by 1 k$\Omega$ for clarity. \textbf{b)} The same as in \textbf{a)} with the opposite polarity of in-plane magnetic field. \textbf{c)} The amplitude of each spin peak as from a fit to each focussing trace. Error bars from the peak fit are smaller than the markers.}
   \label{fig:LargeB}
\end{figure*}

Next, we consider shifts in the Fermi surface caused by large B$_{\parallel}$. Figure~\ref{fig:LargeB} a) shows the evolution of the first focussing peak with a large in-plane magnetic field applied parallel to the QPC current direction (B$_{\parallel}$). As B$_{\parallel}$ increases in magnitude the peak amplitude changes, which would typically interpreted as a change in spin polarisation. Again, when the polarity of B$_{\parallel}$ is reversed (Fig.~\ref{fig:LargeB} b), the change in peak amplitude is not symmetric. This is not consistent with a change in spin polarisation as Zeeman splitting should be the same for $\pm \text{B}_{\parallel}$. To make this clearer, the amplitude of the spin peaks is plotted as a function of B$_{\parallel}$ in Fig.~\ref{fig:LargeB} c). The peak amplitude is clearly not symmetric around B$_{\parallel} = 0$. This amplitude change is consistent across multiple measurements and fitting procedures (e.g. fitting the peak area). See supplementary info for further details \cite{rendell_supp_info}. If the change in peak amplitude was due to Zeeman spin polarisation there should be a monotonic response in the amplitude. This is not visible in Fig.~\ref{fig:LargeB} c), providing further evidence that the change in peak amplitude is not a Zeeman effect or a change in spin polarisation. Furthermore, this amplitude change is not a result of a distortion of the Fermi surface due to B$_{\parallel}$, as this distortion should also be symmetric for $\pm$B$_{\parallel}$.  In addition, the asymmetry in the amplitude change is not an artefact of the setup, as it meets the Onsager reciprocity conditions ($\pm$B$_\parallel$ symmetry is restored if the field direction is reversed and the current and voltage probes are swapped - see supplementary info \cite{rendell_supp_info}).

\begin{figure*}
    \centering
    \includegraphics[width=0.9\textwidth]{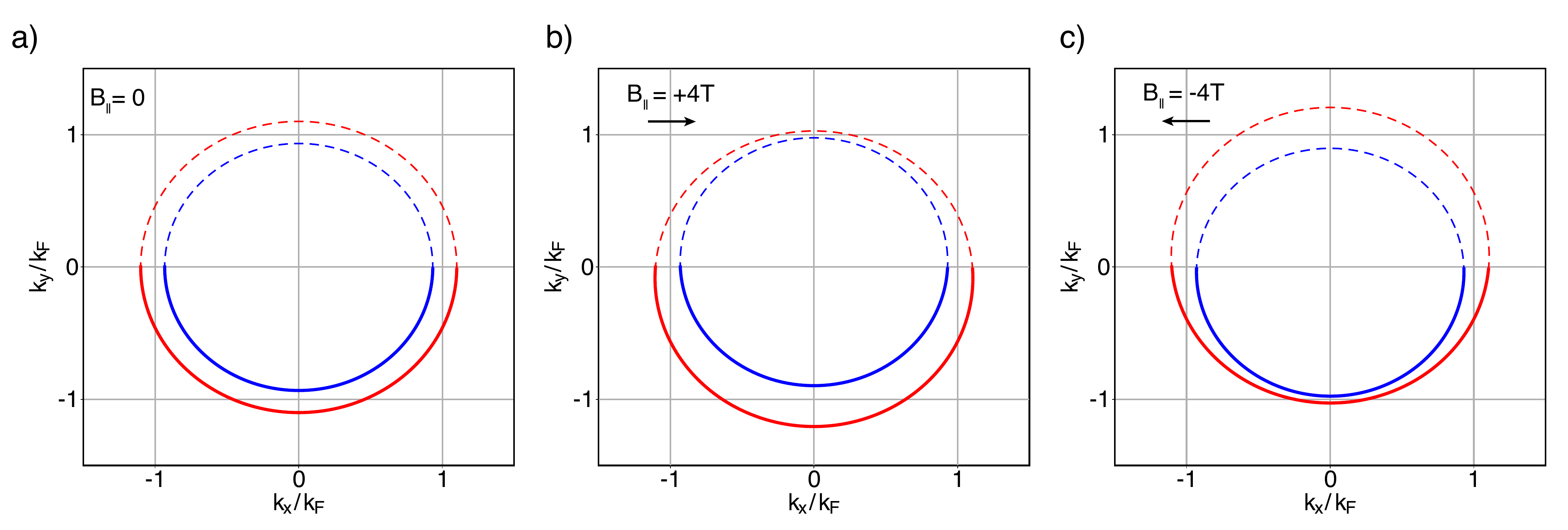}
    \caption{\textbf{Calculated spin split Fermi surface shift with B$_{\parallel} = \pm$ 4T} \textbf{a)} With no B$_{\parallel}$. Red and blue circles correspond to the 2D spin sub-bands. Solid lines indicates the section of the Fermi surfaces over which focussing measurements are performed. \textbf{b)} With B$_{\parallel}$ = +4T, separation of the two spin-split Fermi surfaces is increased over focussing trajectories. \textbf{c)} With B$_{\parallel}$ = -4T, Fermi surfaces almost touch leading to non-adiabatic spin dynamics.}
    \label{fig:NonAdiabatic}
\end{figure*}

The non-monotonic change in the focussing peak amplitude can be explained by a shift in the spin-split Fermi surfaces caused by B$_\parallel$. Figure~\ref{fig:NonAdiabatic} shows the calculated shift in the spin-split Fermi surfaces using Eq.~\ref{eq:Hamiltonian}. Figure~\ref{fig:NonAdiabatic} a) shows the section of the Fermi surface measured by focussing in the absence of B$_{\parallel}$. The red and blue lines indicate the calculated spin-split Fermi surface. The solid section corresponds to the focussing trajectory, while the dashed side does not contribute to focussing. At B$_{\parallel} = 0$T the spacing between the blue and red subbands is the same for all points on the Fermi surface. Figure~\ref{fig:NonAdiabatic} b) shows the change in Fermi surface for B$_{\parallel}$ = +4T. The in-plane field causes the blue Fermi surface to shift towards k$=0$, while the red surface is shifted away. Since only the solid parts of the Fermi surfaces contribute to focussing, these paths move further apart, leading to more adiabatic transport. When the direction of B$_{\parallel}$ is reversed (Fig.~\ref{fig:NonAdiabatic} c), the blue Fermi surface shifts away from k$=0$, while the red surface shifts towards k$=0$. The solid sections of the Fermi surfaces now almost touch, allowing mixing between the spin states. This results in non-adiabatic spin evolution \cite{zener_non-adiabatic_1932} and causes a significant shift in the amplitude and position of the focussing peaks at B$_{\parallel} = -4$T as shown in Fig.~\ref{fig:LargeB} b). This asymmetry in the Fermi surface shift is not observed in other 2D measurements (e.g. Shubnikov-de Haas oscillations) since they sample the full Fermi surface and hence would see the same result for +B$_{\parallel}$ and -B$_{\parallel}$

In summary, we have used magnetic focussing to measure a change in parity of spin-split Fermi surfaces caused by B$_\parallel$. For small B$_\parallel$ the centre of the Fermi surface shifts, resulting in a change in velocity and hence scattering along the two spin resolved focussing trajectories. At large B$_\parallel$, the spin-split Fermi surfaces touch, leading to non-adiabatic transport for one particular direction of the applied B$_\parallel$. This non adiabatic transport causes a significant shift in the focussing peaks. Neither of these effects can be explained by a change in spin polarisation or a Zeeman effect. These results show that magnetic focussing can be used to detect changes in parity of the 2D Fermi surface, as well as its extreme sensitivity to Fermi surface changes. In addition, these effects can only be measured via magnetic focussing as it is able to probe a section of the Fermi surface, unlike other 2D measurements such as Shubnikov-de Haas or commensurability oscillations. While this work has focussed on a change in parity due to the spin-orbit interaction of holes in GaAs, this technique should be applicable to similar changes in the Fermi surface of other 2D systems.

The authors would like to thank U. Z{\"u}licke and Z. Krix for many valuable discussions. Devices were fabricated at the UNSW node of the Australian National Fabrication Facility (ANFF). This research was funded by the Australian Government through the Australian Research Council Discovery Project Scheme; Australian Research Council Centre of Excellence FLEET (project number CE170100039); and by the the UK Engineering and Physical Sciences Research Council (Grant No. EP/R029075/1). All experimental data and calculation code is available at http://dx.doi.org/10.5281/zenodo.8368876

\bibliography{References.bib}

\end{document}